\documentclass[nofootinbib,aps,prb,preprint,notitlepage,superscriptaddress,onecolumn,longbibliography]{revtex4-1}
\usepackage{hyperref}
\hypersetup{colorlinks=true,
		    linkcolor=blue,
		    citecolor=blue,
		    allcolors=blue}
\usepackage{natbib}
\usepackage{amsmath, amsfonts, bm, bbm, braket, mathtools}
\usepackage{graphicx,overpic}
\graphicspath{{./}}

\renewcommand{\H}{\mathcal{H}}  %Hamiltonian
\newcommand{\Z}{\mathbb{Z}}     %Z
\newcommand{\q}{\bm{q}}         %q

\begin{document}

\title{Theory of the field-revealed Kitaev spin liquid}

\author{Jacob S. Gordon}
\author{Andrei Catuneanu}
\affiliation{Department of Physics, University of Toronto, Ontario M5S  1A7, Canada}
\author{Erik S. S\o rensen}
\affiliation{Department of  Physics and Astronomy, McMaster University, Hamilton, Ontario L8S 4M1, Canada}
\author{Hae-Young Kee}
\affiliation{Department of Physics,  University of Toronto, Ontario M5S 1A7, Canada}
\affiliation{Canadian Institute for Advanced Research, CIFAR Program  in Quantum Materials, Toronto, ON M5G 1M1, Canada}
\date{\today}

\maketitle

{\bf
Elementary excitations in highly entangled states such as quantum spin liquids may exhibit exotic statistics, different from those obeyed by fundamental bosons and fermions.
Excitations called non-Abelian anyons are predicted to exist in a Kitaev spin liquid - the ground state of an exactly solvable model proposed by Kitaev almost a decade ago.
A smoking-gun signature of such non-Abelian anyons, namely a half-integer quantized thermal Hall conductivity, was recently reported in $\bm{\alpha}$-RuCl$_3$.
While fascinating, a microscopic theory for this phenomenon in $\bm{\alpha}$-RuCl$_3$ remains elusive because the pure Kitaev phase cannot capture these anyons appearing in an intermediate magnetic field.
Here we present a microscopic theory of the Kitaev spin liquid emerging between the low- and high-field states.
Essential to this result is an antiferromagnetic off-diagonal symmetric interaction that permits the Kitaev spin liquid to protrude from the pure ferromagnetic Kitaev limit under a magnetic field.
This generic model captures a field-revealed Kitaev spin liquid, and displays strong anisotropy of field effects.
A wide regime of non-Abelian anyon Kitaev spin liquid is predicted when the magnetic field is perpendicular to the honeycomb plane.
}

{\it Motivation} -- The Kitaev spin liquid (KSL) is a long-range entangled state on a honeycomb lattice\cite{kitaev2006anyons}, which hosts non-Abelian\cite{kitaev2006anyons,balents2010spin,zhou2017rmp} anyon excitations in a magnetic field.
It has been proposed that topological quantum computation can be performed via braiding of non-Abelian anyons\cite{nayak2008rmp}, meaning the KSL is of both practical, and fundamental interest. 
However, it has been challenging to find a solid state realization of Kitaev physics, which has been the focus of recent research. 
Several honeycomb materials have been suggested as KSL candidates, which are Mott insulators with strong spin-orbit coupling featuring $4d$ or $5d$ transition metal elements\cite{jk2009prl,cjk2010prl,balents2014review,rau2016review,winter2017review}. 
Proposals so far include the iridates A$_2$IrO$_3$\cite{jk2009prl,cjk2010prl,singh2012relevance,modic2014realization,rau2014prl,rau2014trigonal}  (A = Li, Na), and $\alpha$-RuCl$_3$\cite{plumb2014prb,sandilands2015continuum,HSKim2015prb,banerjee2016proximate,sandilands2016excitations}. 
However, all these candidates exhibit magnetic ordering at low temperatures\cite{choi2012prl,cjk2014zigzag,fletcher1967magnetic,sears2015prb,johnson2015monoclinic,banerjee2016proximate,cao2016structure,HSKim2016structure,janssen2017model}, which masks potential Kitaev physics. 
Later theoretical\cite{yadav2016field,lampenkelley2018induced,liu2018dirac} and experimental\cite{baek2017evidence,wolter2017induced,zheng2017gapless} results suggest that $\alpha$-RuCl$_3$ may enter a field-induced spin liquid, but there has been no evidence that it is a chiral spin liquid until a half-integer quantized thermal Hall conductivity was reported in $\alpha$-RuCl$_3$\cite{kasahara2018thermal}; a strong indication\cite{vinkleraviv2018approximate,cookmeyer2018spinwave,ye2018quantization} of chiral edge currents of Majorana fermions (MFs) predicted in a KSL.

While this is a first experimental evidence of charge-neutral non-Abelian anyons in spin systems, a microscopic theory describing the appearance of such a chiral KSL under a field in $\alpha$-RuCl$_3$ is missing.
This is because if the dominant interaction in $\alpha$-RuCl$_3$ is the ferromagnetic (FM) Kitaev term -- based on spin wave analysis\cite{cookmeyer2018spinwave} and ab-initio studies\cite{HSKim2016structure,janssen2017model} -- the FM Kitaev phase is almost immediately destroyed, and the polarized state appears in an applied field\cite{jiang2011JK,fu2018robust,liang2018intermediate} with no intervening phase.
This can be contrasted with the antiferromagnetic (AFM) Kitaev phase which hosts a potentially gapless spin liquid under a field, supported by several numerical studies\cite{fu2018robust,gohlke2018dynamical,liang2018intermediate,nasu2018successive,hickey2018visons,ronquillo2018orientation,lu2018spinon,zou2018neutral,patel2018spinon}.
However, this intermediate gapless U(1) spin liquid cannot explain the half-integer thermal Hall effect observed in $\alpha$-RuCl$_3$, even though it is an intriguing spin liquid.
Thus searching for a possible spin liquid with MFs leading to the half-integer thermal Hall effect under a magnetic field remains a challenging task. 

Here we present a microscopic theory of the KSL displaying a half-integer thermal Hall effect under a magnetic field.
The key to our result is an AFM symmetric off-diagonal $\Gamma$ interaction, essential to stabilize the KSL for intermediate fields.
The KSL emerges between the low- and high-field phases as $\Gamma$ increases, and is connected to the pure FM Kitaev phase at zero field.
We introduce the microscopic theory with a brief review of the generic nearest neighbour spin model for spin-orbit coupled honeycomb materials, appropriate for $\alpha$-RuCl$_3$.

\begin{figure}
\includegraphics[width=0.9\linewidth,trim={0mm 0mm 110mm 0mm},clip]{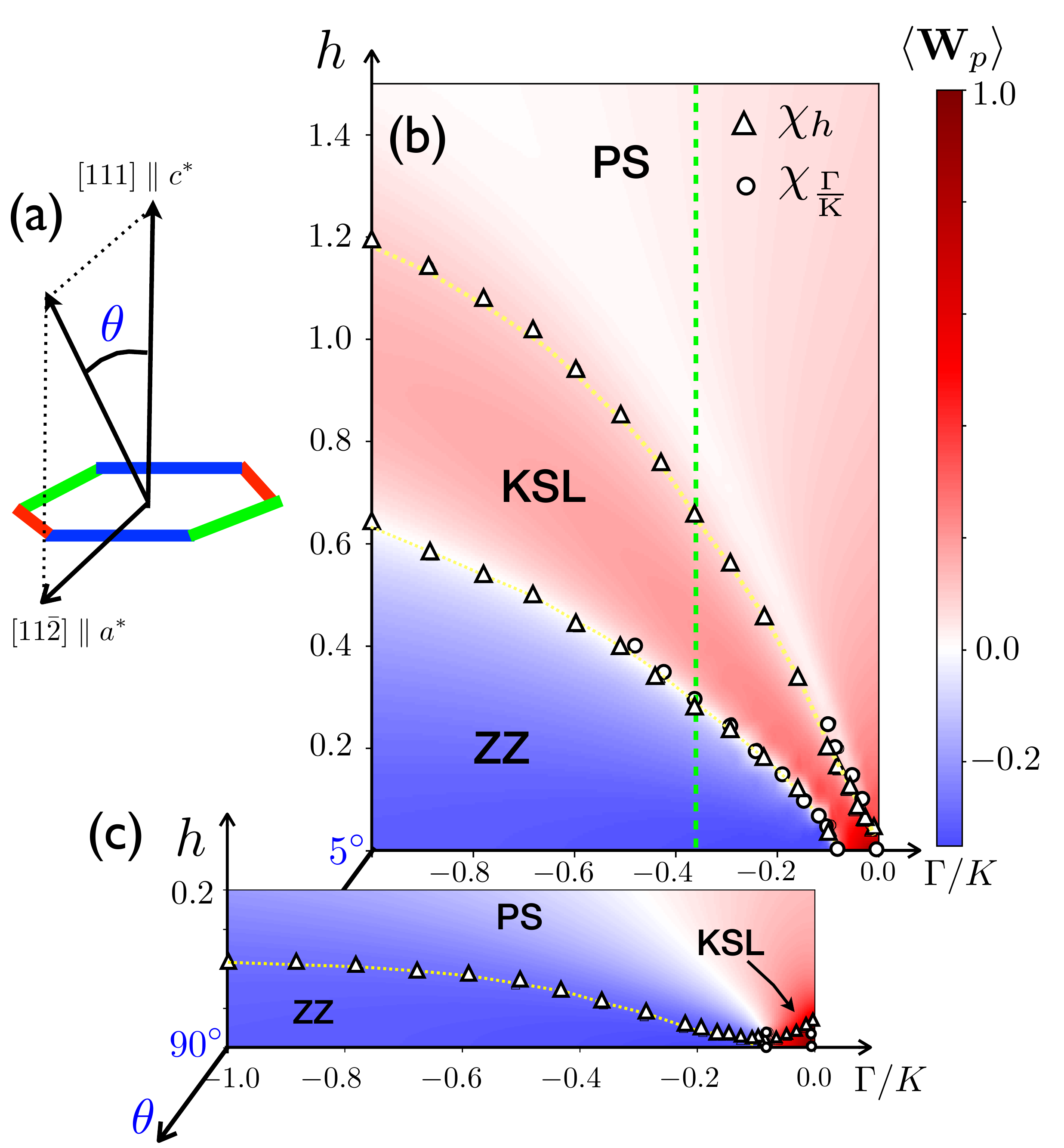}
\caption{
(a) The angle $\theta$ is measured from $[111]$ towards the in-plane $[11\overline{2}]$ direction.
Phase diagrams in the $\Gamma/K$--$h$ plane (FM Kitaev and AFM $\Gamma$) obtained from the 24-site ED honeycomb cluster are shown for (b) $\theta = 5^{\circ}$ and (c) $\theta = 90^{\circ}$ at a fixed $\Gamma^{\prime} = -0.03$ in units of $\sqrt{K^2 + \Gamma^2} \equiv 1$.
Circular and triangular symbols represent peaks in the susceptibilities $\chi_{\scriptscriptstyle \Gamma/K}$ and  $\chi_{h}$, respectively. 
The intermediate-field KSL is seen to be adiabatically connected to the pure $K$ limit.
Colours represent the expectation value of the plaquette operator $\braket{W_p}$, which is discussed further in the main text.
A sequence of phase transitions from ZZ order to the KSL, and finally the PS is found for $\theta = 5^{\circ}$, except near the pure $K$ limit. 
The green line in (b) at $\Gamma/K \simeq -0.37$ denotes a representative slice where $\chi_h$ and $S(\bm{q})$ are plotted in Fig.~\ref{fig:C3-SSF-Xh}.
}
\label{fig:C3-phase-diagram}
\end{figure}

{\it Model} -- The nearest neighbour model has been derived in [\onlinecite{jk2009prl},\onlinecite{rau2014prl},\onlinecite{rau2014trigonal}] based on a strong coupling expansion of the Kanamori Hamiltonian.
The combination of crystal field splitting and strong spin-orbit coupling leads to a model based on pseudospin-$\tfrac{1}{2}$ local moments with bond-dependent interactions.
On a bond of type $\gamma\in\{x,y,z\}$ with sites $j,k$, the nearest-neighbour spin Hamiltonian is taken to be of the $J$-$K$-$\Gamma$-$\Gamma^{\prime}$ form\cite{rau2014trigonal}
\begin{align}\label{eq:hamiltonian}
\begin{split}
  \H_{jk}^{\gamma} &= J \bm{S}_j  \cdot \bm{S}_k + K S_j^{\gamma}S_k^{\gamma} + \Gamma (S_j^{\alpha}S_k^{\beta} + S_j^{\beta}S_k^{\alpha}) \\
                   &\ \ + \Gamma^{\prime}(S_j^{\alpha}S_k^{\gamma} + S_j^{\gamma}S_k^{\alpha} + S_j^{\beta}S_k^{\gamma} + S_j^{\gamma}S_k^{\beta})
\end{split}
\end{align}
where $\alpha,\beta$ are the remaining spin components in $\{x,y,z\}\setminus\{\gamma\}$.
The spin components are directed along the cubic axes of the underlying ligand octahedra, so the honeycomb layer lies in a plane perpendicular to the [111] spin direction as shown in Fig.~\ref{fig:C3-phase-diagram}(a).
A small $\Gamma^{\prime}$ is present due to trigonal distortion of ligand octahedra in the real material.
Here we omit the Heisenberg $J$  for simplicity, and its effects are discussed later.
Earlier studies\cite{plumb2014prb,janssen2017model,catuneanu2018path,yb2018signatures,liu2018dirac} noted that the $\Gamma$ interaction with AFM sign may play an important role near the FM Kitaev regime to stabilize the spin liquid\cite{catuneanu2018path}.
Since $\alpha$-RuCl$_3$ has a dominant FM Kitaev interaction with AFM $\Gamma$, we focus on $\Gamma/K\in[-1,0]$ with $\Gamma > 0$ and $K < 0$.
The remaining parameters of the Hamiltonian are expressed in units of $\sqrt{K^2 + \Gamma^2} \equiv 1$.

To describe the effect of a magnetic field, we consider a Zeeman term with isotropic $g$-factor
\begin{equation}\label{eq:zeeman}
  \H_Z = -h\sum_{j}\bm{\hat{h}}\cdot\bm{S}_j,
\end{equation}
where $h$ is the magnetic field strength, and $\bm{\hat{h}}$ is a unit vector specifying the field direction.
In order to make a connection with the thermal Hall measurements\cite{kasahara2018thermal}, we focus on field directions in the $\bm{\hat{a}}\bm{\hat{c}^*}$ plane spanned by $[11\overline{2}]$ and $[111]$.
The direction of the field in this plane is parameterized by an angle $\theta$ from the $[111]$ direction, as shown in Fig.~\ref{fig:C3-phase-diagram}(a).

{\it ED Results} -- Our main results are shown in Fig.~\ref{fig:C3-phase-diagram}.
Phase diagrams in the $\Gamma/K$ -- $h$ plane are shown for tilting angles $\theta = 5^{\circ}$ and $90^{\circ}$ obtained through numerical exact diagonalization (ED).
Details of the 24-site honeycomb cluster used are discussed in the Supplementary Information (SI).
Peaks in the susceptibilities $\chi_{\scriptscriptstyle \Gamma/K} = -\partial^2_{\scriptscriptstyle \Gamma/K}e_0$ and $\chi_h = -\partial^2_{h}e_0$, where $e_0 = E_0/N$ is the ground state energy density, are depicted as triangles and circles, respectively. 
There are three phases in the phase space, namely, ZZ magnetic order at low fields, the KSL, and a polarized state (PS) at high fields.
Remarkably, we find {\it an intermediate KSL} sitting between ZZ order and the PS, which is adiabatically connected to the pure $K$ limit.

\begin{figure}
\includegraphics[width=1\linewidth]{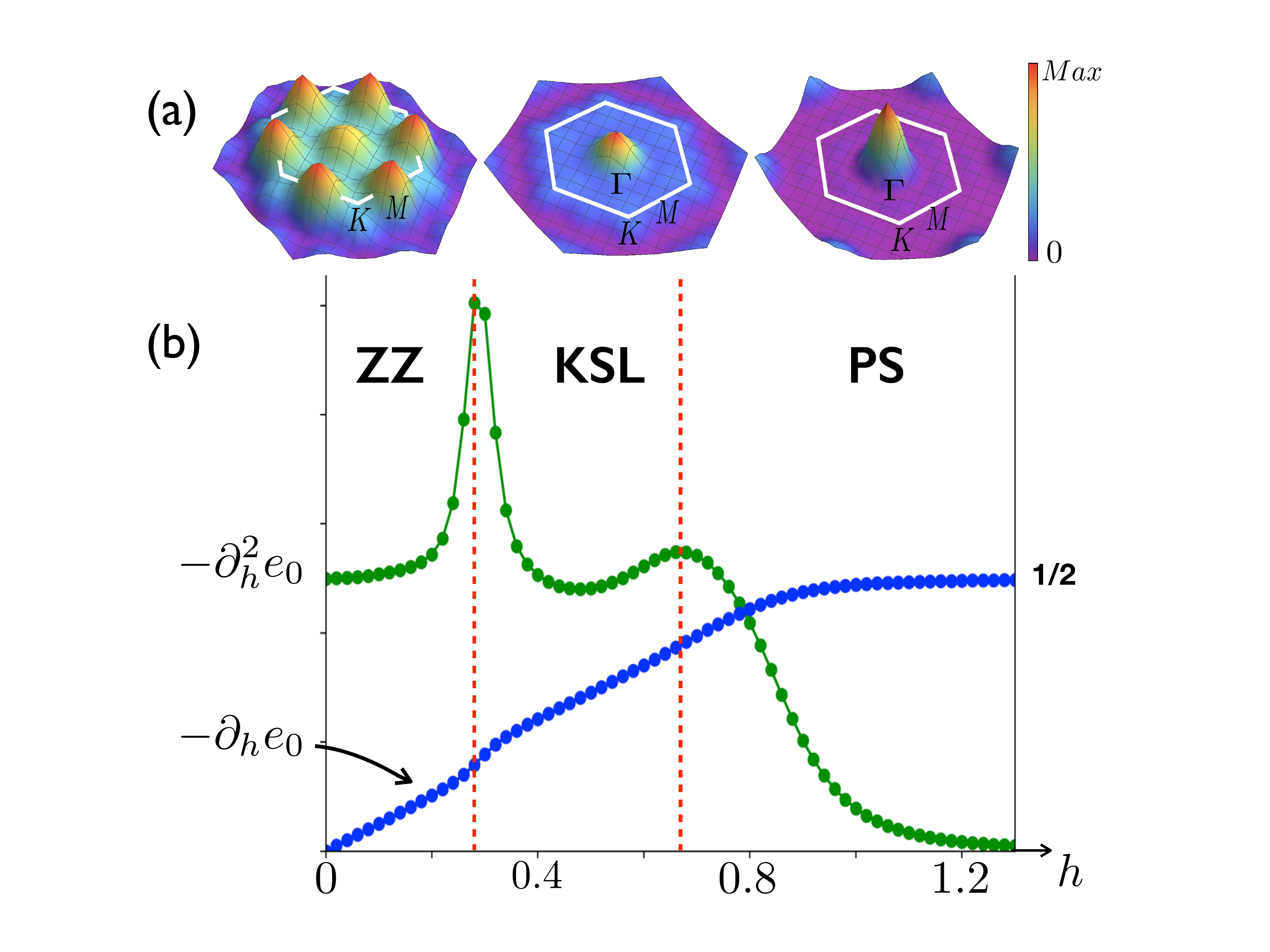}
\caption{
(a) Spin structure factor $S(\bm{q})$ within the ZZ, KSL, and PS phases.
(b) $\chi_h= -\partial^2_{h}e_0$ and magnetization $m= -\partial_{h}e_0$ as a function of a $5^{\circ}$ tilted field are shown for a fixed $\Gamma/K \simeq -0.37$, indicated by a green dashed line in Fig.~\ref{fig:C3-phase-diagram}(b).
}
\label{fig:C3-SSF-Xh}
\end{figure}

The intermediate KSL begins from the pure FM $K$ regime, which is unstable to a small magnetic field in the $[111]$ direction.
However, it is stabilized by the AFM $\Gamma$ term, and extends above the ZZ phase in a magnetic field.
For moderate $\Gamma/|K|$ appropriate for $\alpha$-RuCl$_3$, we observe a sequence of phase transitions from ZZ order to the KSL, and finally to the PS as shown in Fig.~\ref{fig:C3-SSF-Xh}(b) for $\Gamma/K \simeq -0.37$.
Interestingly, with a constant $\Gamma^{\prime}$ both the ZZ and KSL phase spaces widen with increasing $\Gamma$, as suggested by the curvature of the transition line in Fig.~\ref{fig:C3-phase-diagram}(b).
This behaviour survives for further tilting of the magnetic field away from $[111]$, with increased $[11\overline{2}]$ in-plane component.
However, the window of the KSL rapidly diminishes with tilting angle until a direct transition between ZZ order and the PS appears at much smaller field, as shown in Fig.~\ref{fig:C3-phase-diagram}(c) for a $[11\overline{2}]$ field.
The critical field required to destroy the ZZ ordering drops drastically with increasing $\theta$.
With an estimate of the energy unit as $\sqrt{K^2 + \Gamma^2} \sim 7\ \mathrm{meV}$, $h = 0.1$ corresponds to a field of $\sim10$ T.
This is within the range of fields required to kill the ZZ order in $\alpha$-RuCl$_3$\cite{kasahara2018thermal}.

Since the pure Kitaev limit involves the fractionalization of spins into itinerant MFs and $\Z_2$ fluxes, another quantity that characterizes the KSL is the plaquette operator $W_p$\cite{kitaev2006anyons}:
\begin{equation}
W_p = 2^6 S_1^xS_2^yS_3^zS_4^xS_5^yS_6^z,
\end{equation}
defined on sites belonging to a hexagonal plaquette $p$.
The pure KSL with $h = \Gamma = \Gamma^{\prime} = 0$ (bottom right corner of the phase diagram) is a flux-free state with $\braket{W_p}  = +1$ on all plaquettes.
A finite $\Gamma$, $\Gamma^{\prime}$ or $h$ spoils the exact solubility of the Kitaev model, as they generate interactions among the MFs and $\Z_2$ fluxes.
Although the plaquette operators are no longer conserved quantities, $\braket{W_p}$ remains positive and can distinguish the KSL from the neighbouring ZZ ordered phase.
The ZZ and PS phases have finite plaquette expectation and plaquette-plaquette correlations due to the magnetic order.
In particular, a FM product state in the polarized phase at high fields has positive $\braket{W_p}$ for small $\theta$, which eventually vanishes unless the moment involves all three spin components.
Conversely, ZZ magnetic order has negative plaquette expectation value.
Further details of this calculation can be found in the SI.
The plaquette expectation is negative in the ZZ phase, and positive in the KSL, so $\braket{W_p}$ should vanish at the ZZ-KSL phase transition.
This behaviour of $\braket{W_p}$ is seen with ED on the 24-site honeycomb cluster in Fig.~\ref{fig:C3-phase-diagram}.
It is interesting to note that there is no sharp change in $\braket{W_p}$ between the ZZ and PS when the field is along $[11\overline{2}]$, while a remnant of the vanishing $\braket{W_p}$ is found at higher fields.

To confirm the ZZ magnetically ordered phase at low field, we compute the spin structure factor $S(\q)$, which displays sharp features at the $\bm{M}$-point of the Brillouin zone (BZ) as shown in Fig.~\ref{fig:C3-SSF-Xh}(a).
Within the KSL, $S(\q)$ is diffuse, with a soft peak at the $\bm{\Gamma}$-point.
As expected, the PS exhibits a sharp feature at the $\bm{\Gamma}$-point.
At large fields, the magnetization $m = -\partial_h e_0$ saturates within the PS as shown in Fig.~\ref{fig:C3-SSF-Xh}(b).
ZZ ordering at low field can be traced back to the presence of other small interactions, in this case due to $\Gamma^{\prime}$\cite{jk2009prl,rau2014prl,rau2014trigonal,yadav2016field}.
At zero field with $\Gamma^\prime = -0.03$ only, the ZZ order disappears at $\Gamma/K \simeq -0.09$ in the 24-site cluster leaving a small window of the KSL intact.
A larger $\Gamma^{\prime}$ enlarges the ZZ ordered phase, and decreases the window of KSL at $h=0$ as shown in Fig. 2 of the SI.

{\it DMRG Results} -- Due to fundamental limitations on system sizes accessible with ED, we have also studied a two-leg honeycomb strip using density-matrix renormalization group (DMRG), which can access system sizes an order of magnitude larger.
We denote the total number of sites in the strip by $N$.
This geometry has recently been used to study the Kitaev-Heisenberg model\cite{catuneanu2018ladder}, where it was found that its phase diagram displays a striking similarity with that of the 2D honeycomb lattice.
Further rationale for this choice of geometry is discussed in the SI.

\begin{figure}
\includegraphics[width=0.8\linewidth,trim={0mm 0mm 160mm 0mm},clip]{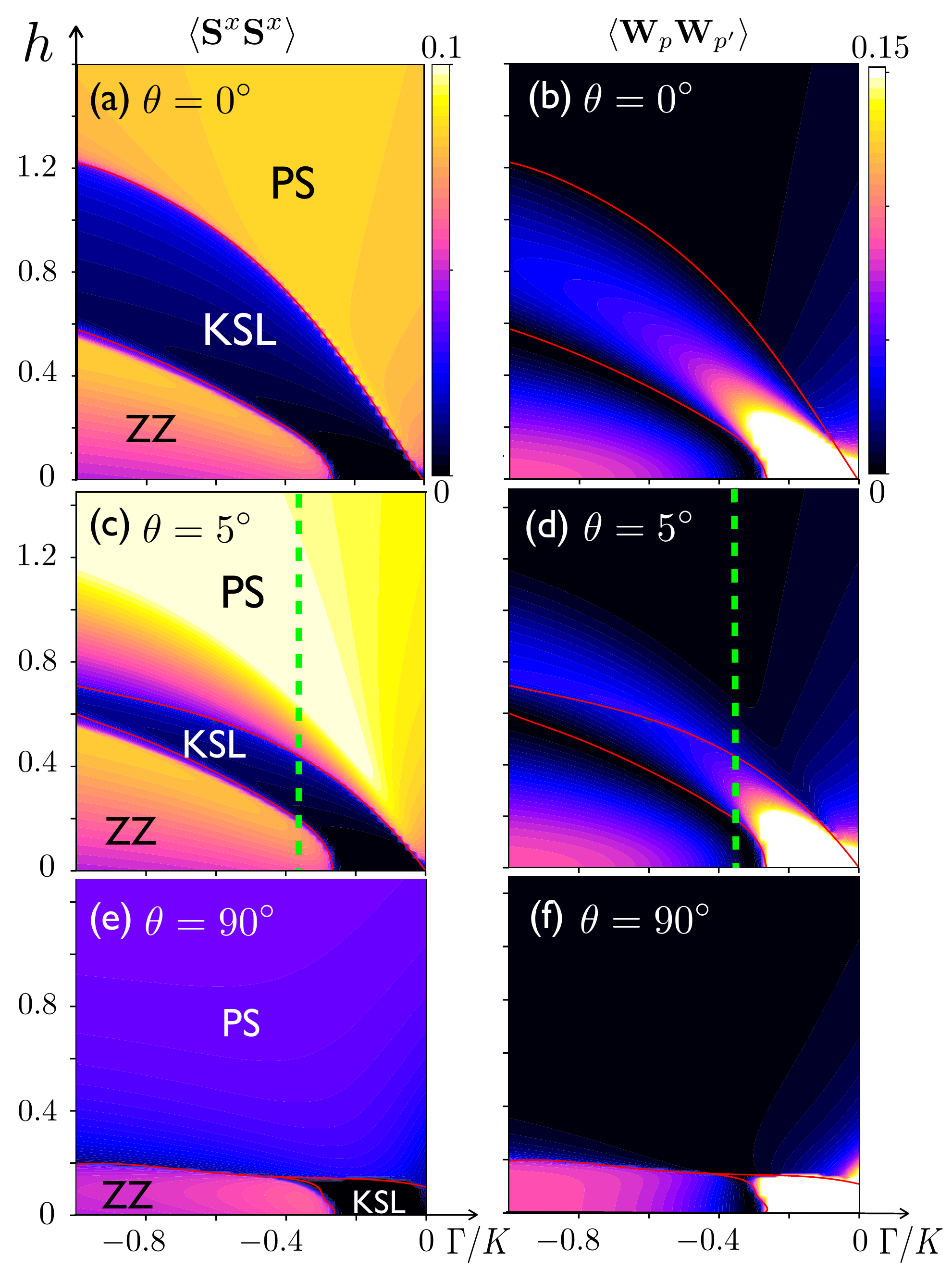}
\caption{
The spin-spin correlator $\braket{S_j^{x}S_k^{x}}$ at $k-j = 50$ along one leg of the strip is shown in the first column, and the plaquette-plaquette correlator $\braket{W_p W_{p'}}$ at $p'-p = 30$ in the second as a function of $\Gamma/K$ and $h$ with $\Gamma^{\prime} = -0.1$.
These are obtained in the two-leg honeycomb strip with DMRG with $N=200$ and OBC.
The field directions are (a-b) $\theta = 0^{\circ}$, (c-d) $\theta = 5^{\circ}$, and (e-f) $\theta = 90^{\circ}$ ([11$\overline{2}$] in-plane).
Smooth curves fitted to peaks in either $\chi_h$, or $\chi_{\scriptstyle \Gamma/K}$ are drawn with red lines.
The dashed green line at $\Gamma/K = -0.325$ indicates a representative slice where certain quantities are plotted in Fig.~\ref{fig:dmrg-slice} for a $\theta = 5^{\circ}$ tilted field.
}
\label{fig:dmrg-wpwp-xx}
\end{figure}

Phase diagrams in the $\Gamma/K$ -- $h$ plane with $N = 200$ and open boundary conditions (OBC) for tilting angles $\theta=0^\circ$, $5^\circ$ and $90^\circ$ are shown in Fig.~\ref{fig:dmrg-wpwp-xx}.
The phase boundaries in Fig.~\ref{fig:dmrg-wpwp-xx}, determined by peaks in $\chi_h$ or $\chi_{\scriptscriptstyle \Gamma/K}$, are represented by red lines. 
We find a notable similarity with the ED phase diagram of Fig.~\ref{fig:C3-phase-diagram}, showing a large region of KSL which extends above the ZZ ordered phase and below the PS under a magnetic field.
As the in-plane component of the field becomes larger, the intermediate KSL phase space rapidly shrinks.
As we found with ED on the 24-site honeycomb cluster, the intermediate KSL at large $\Gamma/|K|$ disappears, leaving a single direct transition from ZZ to the PS as the field tilts towards $90^\circ$.
Crucially, a small region of KSL remains intact at smaller $\Gamma/|K|$.
Thus, with the in-plane $[11\overline{2}]$ field, the KSL is confined to a narrow range of field near the pure FM Kitaev limit.
This constrasts with another in-plane field direction $[1\overline{1}0]$, shown in the SI, where the KSL is immediately destroyed by any non-zero field for $\Gamma/|K| < \Gamma/|K|_c$, where $\Gamma/|K|_c$ refers to the transition point between the ZZ and KSL at $h=0$. 
On the other hand, for any $\Gamma/|K| > \Gamma/|K|_c$, there is a direct transition between the ZZ and the PS at finite field.
This is consistent with the observation that there is no $\mathbb{Z}_2$ topological order when a certain pseudo-mirror symmetry is preserved\cite{zou2018neutral}, as our Hamiltonian with a $[1\overline{1}0]$ field direction has this symmetry.

Figures~\ref{fig:dmrg-wpwp-xx}(a), (c), and (e) show $\braket{S^x_jS^x_k}$ at separation $k - j = 50$ along one leg of the strip as a function of field and $\Gamma/K$ for different tilting angles of the field in the $\bm{\hat{a}}\bm{\hat{c}^*}$ plane.
As expected, correlations are appreciable within the magnetically ordered and polarized states.
The KSL phase is very clearly distinguished from the surrounding ordered states by nearly vanishing $\braket{S^xS^x}$ ($= \braket{S^yS^y}$) spin correlations.
However, spin-spin correlations need not be identically zero except at $h = 0$ due to a component of the spin 
aligning with the field, which is more pronounced when $h$ is large.

In Fig.~\ref{fig:dmrg-wpwp-xx}(b), (d) and (f) we show plaquette-plaquette correlations $\braket{W_pW_{p'}}$ at separation $p'- p = 30$.
Close to the Kitaev limit these correlations are nearly unity, consistent with $\braket{W_p} = +1$ in the $K$ limit, and decrease with increasing field and $\Gamma/|K|$ within the KSL.
This is expected because the magnetic field, as well as $\Gamma,\Gamma^{\prime}$, introduces interactions among the MF and flux degrees of freedom.
Interestingly, the plaquette-plaquette correlations, which approach $|\braket{W_p}|^2$ in the KSL at large separations, show large fluctuations above the KSL phase.
This effect can also can be seen with ED on the 24-site honeycomb cluster shown in Fig.~\ref{fig:C3-phase-diagram}(b), as noted by remaining $\braket{W_p}$ variations above the transition line. 

\begin{figure}
\includegraphics[width=\linewidth,trim={10mm 20mm 55mm 5mm},clip]{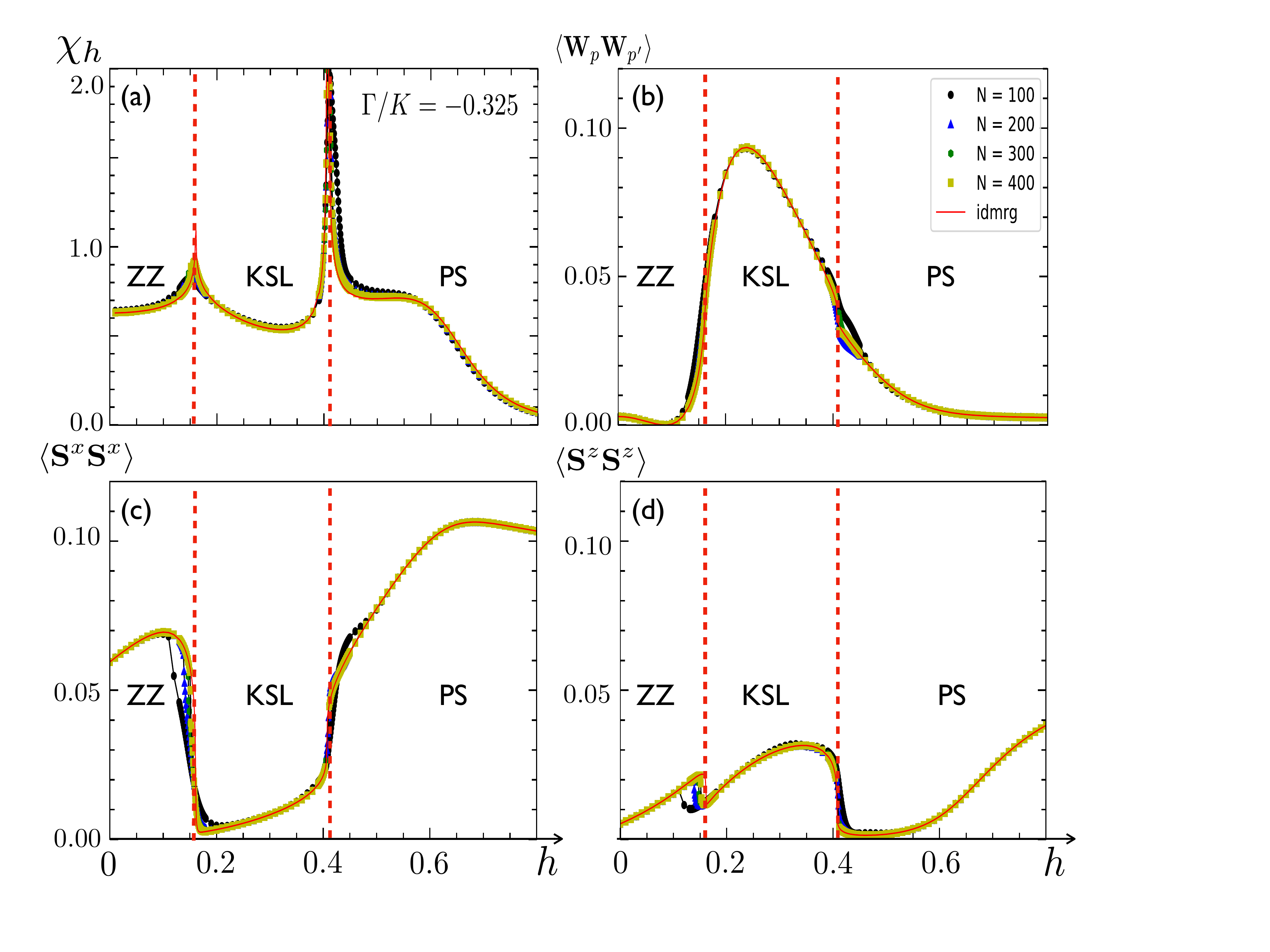} %trim {L, B, R, T}
\caption{
With $\Gamma/K = -0.325$ and $\Gamma^{\prime} = -0.1$, quantities computed in the two-leg honeycomb strip with DMRG are plotted as a function of a $5^{\circ}$ field for system sizes $N = 100, 200, 300, 400$ and infinite DMRG (iDMRG) as solid red line. 
(a) $\chi_h$ displays two peaks which sharpen with increasing system size.
These transitions separate ZZ magnetic order and the PS from the intervening KSL.
(b) Plaquette-plaquette correlations $\braket{W_pW_{p'}}$ are small within the ZZ and polarized phases, and become appreciable within the KSL.
(c) Conversely, the $xx$ spin-spin correlations $\braket{S^xS^x}$ are small within the KSL phase, and large in the magnetically ordered phases.
(d) Sharp changes in the $zz$ spin-spin correlations are seen across the phase boundaries, and display a finite value consistent with a KSL in field.
Spin and plaquette correlations are evaluated at the largest separation accessible in each cluster.
}
\label{fig:dmrg-slice}
\end{figure}

A representative cut of the phase diagram is presented in Fig.~\ref{fig:dmrg-slice}(c-d) as a function of a $5^{\circ}$ tilted field with $\Gamma/K = -0.325$, which corresponds to the green line in Fig.~\ref{fig:dmrg-wpwp-xx}(c) and (d).
With increasing field, a sequence of transitions from ZZ order to the KSL and finally the PS are evidenced by strong singular behaviour of $\chi_h$ in Fig.~\ref{fig:dmrg-slice}(a).
The transition between ZZ order and the KSL is accompanied by a sharp increase in plaquette-plaquette correlations shown in Fig.~\ref{fig:dmrg-slice}(b), and a larger value of $\braket{W_p}$ accordingly.
Components $\braket{S^xS^x}$ and $\braket{S^zS^z}$ of the spin-spin correlators are plotted in Fig.~\ref{fig:dmrg-slice}(c) and (d).
While the $\braket{S^xS^x}$ correlations are small in the KSL, the $\braket{S^zS^z}$ correlations are slightly larger.
This is similar to the honeycomb cluster with ED, where a finite magnetization $m$ is present in the KSL phase, as shown in Fig.~\ref{fig:C3-SSF-Xh}(b).
Asymmetry between the $S^x$ and $S^z$ components of the spin is due to the two-leg honeycomb strip connectivity, and tilting of the magnetic field.
The preceding properties are shown for $N = 100, 200, 300, 400$ and iDMRG in Fig.~\ref{fig:dmrg-slice} with different colours, and are seen to be relatively insensitive to the system size.

\begin{figure}
\includegraphics[width=\linewidth,trim={0mm 0mm 75mm 0mm},clip]{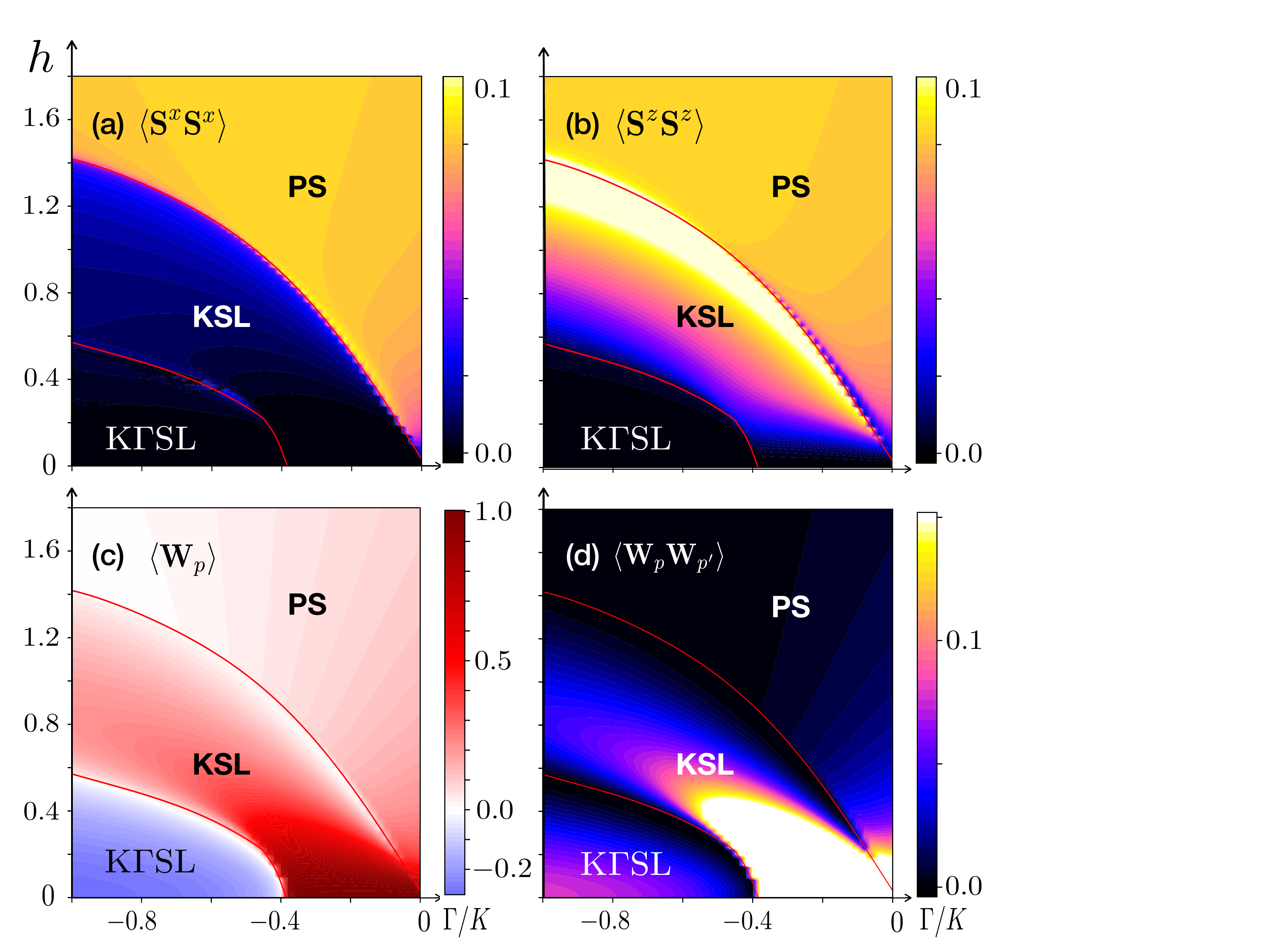} %trim {L, B, R, T}
\caption{
Underlying phase diagram with $\Gamma^{\prime} = 0$ and $\theta=0$ in the $\Gamma/K$ -- $h$ plane as calculated with DRMG for $N=200$ and OBC.
Spin-spin correlations (a) $\braket{S^x_jS^x_k}$ and (b) $\braket{S^z_jS^z_k}$ at $k-j=50$ are absent within the K$\Gamma$SL and the KSL phases at $h = 0$, and develop under the field.
Phase boundaries, determined by smoothed fits to the peaks of of $\chi_h$ and $\chi_{\scriptstyle \Gamma/K}$, are drawn as red lines, and separate the K$\Gamma$SL and the PS from the intervening KSL.
(c) Plaquette expectation $\braket{W_p}$ differentiates the K$\Gamma$SL and the KSL, as it changes sign across the transition. 
(d) Plaquette-plaquette correlations $\braket{W_pW_{p'}}$ at $p'-p=30$ are small in the PS and become appreciable within the K$\Gamma$SL and KSL, approaching $\braket{W_p}^2$ in the limit of large separation.
}
\label{fig:kgsl}
\end{figure}

{\it Underlying Phase Diagram} --
To understand the microscopic mechanism of the emerging KSL, we study the $K$--$\Gamma$ model without $\Gamma^{\prime}$ at $\theta = 0$. 
At zero field, there is a finite region of the KSL when the AFM off-diagonal symmetric $\Gamma$ interaction is introduced.
In the absence of $\Gamma^{\prime}$, there is a transition in zero field at $\Gamma/K \simeq -0.4$ from the KSL to another spin liquid dubbed $\mathrm{K}\Gamma$ spin liquid (K$\Gamma$SL)\cite{lampenkelley2018induced}.
Components of the spin-spin correlators, $\braket{S^xS^x}$ and $\braket{S^zS^z}$ shown in Fig.~\ref{fig:kgsl}(a-b), demonstrate a lack of magnetic order in the KSL and K$\Gamma$SL at $h=0$, and finite correlations building with increasing field.

The K$\Gamma$SL is characterized by a finite $\braket{W_pW_{p'}}$ like the KSL, but with {\it negative} $\braket{W_p}$ as shown in Fig.~\ref{fig:kgsl}(c-d).
While $\braket{W_p}$ is positive in the KSL, a negative $\braket{W_p}$ in the K$\Gamma$SL indicates a novel phase with a finite flux density.
Strikingly, when the field is applied along the $[111]$ direction, the KSL sits above the K$\Gamma$SL for a fixed $\Gamma/K$ leading to two phase transitions with increasing field: K$\Gamma$SL $\rightarrow$ KSL $\rightarrow$ PS.
The K$\Gamma$SL is extremely fragile to additional interactions that stabilize ZZ order.
For example, when a small $\Gamma^{\prime}$ interaction is introduced, the K$\Gamma$SL is replaced by the ZZ ordered phase as shown in Fig.~\ref{fig:C3-phase-diagram}.
Importantly, the ZZ order does not extend all the way to the Kitaev limit, and leaves a finite region of the KSL at zero field.

{\it Discussion} -- In $\alpha$-RuCl$_3$ it was suggested that the FM nearest $(-J)$ and AFM third neighbour ($J_3$) Heisenberg interactions are important to stabilize the ZZ magnetic order\cite{winter2016challenges}.
Indeed the combination of $-J$ and $J_3$ has a similar effect to $\Gamma^{\prime}$ in that they both induce ZZ order.
Thus, their inclusion would not alter our main results.
However, if the combined strength of $\Gamma^{\prime}$, $-J$ and $J_3$ is too large, it would completely wipe out the intermediate KSL.
Indeed we show that a larger $\Gamma^{\prime}$ causes ZZ order to overtake the KSL in Fig. 2 of the SI.
Experimental reports of a half-quantized thermal Hall conductivity in $\alpha$-RuCl$_3$ imply that the combined strength of these parameters is small enough to leave the intermediate KSL intact, yet finite to induce the ZZ order.
It is possible that the K$\Gamma$SL survives with smaller $\Gamma^{\prime}$ while developing ZZ order, resulting in two spin liquids between ZZ and PS under a field.
Quantifying these strengths is left for a detailed future study.

As the magnetic field is tilted away from the out-of-plane $[111]$ direction towards the in-plane $[11\overline{2}]$ direction the intermediate KSL region shrinks rapidly,
 independent of the strength of $\Gamma^{\prime}$, and for both cluster shapes studied here.
What remains is a small intermediate phase at fields an order of magnitude smaller for moderate $\Gamma/|K|$, showing a dramatic magnetic anisotropy.
While smaller tilting angles are less effective at destroying the ZZ magnetic order, they offer a much larger region of the KSL.
To enlarge the intermediate KSL phase, we therefore propose that a field should be applied at smaller tilting angles.
Furthermore, for another in-plane field direction $[1\overline{1}0]$, the phase diagram exhibits no intermediate-field KSL at any finite $\Gamma/|K|$ or $h$ as discussed earlier.
Further thermal Hall transport measurements for different in-plane directions would be desirable in order to test our microscopic theory. 

There are fascinating aspects of this work that require further study.
The first is the presence of large fluctuations of $\braket{W_pW_{p^{\prime}}}$ and $\braket{S_j^xS_k^x}$ just above the KSL phase into the PS, which is also seen by $\chi_{\scriptscriptstyle \Gamma/K}$ and an unsaturated magnetization in the 24-site honeycomb cluster.
This is suggestive of a non-trivial crossover region into the polarized phase.
We also note the presence of a novel phase dubbed K$\Gamma$SL in the underlying phase diagram of the $K$-$\Gamma$ model next to the KSL phase.
The K$\Gamma$SL phase is differentiated from the KSL by a sharp drop from $\braket{W_p} = 1$ in the KSL to $\braket{W_p} \simeq -0.35$ in the K$\Gamma$SL, accompanied by a singular $\chi_{\scriptscriptstyle \Gamma/K}$.
Nature of the K$\Gamma$SL, and the transition to the KSL are excellent subjects for future study.
For instance, possible vortex patterns due to strong interactions among MFs and $\Z_2$ vortices would be highly interesting to pursue.

\bibliography{references}
\vspace{2mm}

\noindent {\bf Supplementary Information} is available in the online version of the paper. \\[-0.5mm]

\noindent {\bf Acknowledgements} 
This work was supported by the Natural Sciences and Engineering Research Council of Canada and the Center for Quantum Materials at the University of Toronto. 
This research was enabled in part by support provided by Sharcnet (\href{http://www.sharcnet.ca}{www.sharcnet.ca}) and Compute Canada (\href{http://www.computecanada.ca}{www.computecanada.ca}).
Computations were performed on the GPC and Niagara supercomputers at the SciNet HPC Consortium. 
SciNet is funded by: the Canada Foundation for Innovation under the auspices of Compute Canada; the Government of Ontario; Ontario Research Fund - Research Excellence; and the University of Toronto. \\[-0.5mm]

\noindent {\bf Author Contributions}
Exact diagonalization calculations performed by J.S.G. and A.C.
Density-matrix renormalization group calculations performed by E.S.S.
H.-Y.K. planned and supervised the project.
All authors wrote the manuscript. \\[-0.5mm]

\noindent {\bf Author Information}
The authors declare no competing financial interests. Correspondence should be addressed to H.-Y.K. (\href{mailto:hykee@physics.utoronto.ca}{hykee@physics.utoronto.ca}). \\[-0.5mm]

\noindent {\bf Methods \label{sec:methods}} 
Numerical exact diagonalization (ED), and density matrix renormalization group (DMRG) are used to study the parameter space appropriate for $\alpha$-RuCl$_3$. 
In the ED and DMRG calculations, we consider the two-leg honeycomb strip geometry, and the 24-site honeycomb shape in ED only, both shown in Fig. 1 of the Supplementary Information (SI).  
The choice of these clusters is discussed in the SI, and is related to hidden points of SU(2) symmetry present in the 2D limit.

ED was performed on the 24-site honeycomb cluster with periodic boundary conditions, where the Lanczos method was used to obtain the lowest-lying eigenvalues and eigenvectors of the Hamiltonian in Eq.~(\ref{eq:hamiltonian}).

Part of the numerical calculations were performed using the ITensor library (\href{http://itensor.org}{http://itensor.org}) typically with a target precision of $10^{-11}$ using up to $1{,}000$ states.
All DMRG calculations were performed using open boundary conditions (OBC).
The iDMRG calculations were performed using a target precision of $5\times 10^{-11}$ and up to $1{,}000$ states.

\end{document}